\documentclass{egpubl}
\usepackage{egsgp18}
\usepackage{amssymb}

\SpecialIssuePaper         
 \electronicVersion 

\ifpdf \usepackage[pdftex]{graphicx} \pdfcompresslevel=9
\else \usepackage[dvips]{graphicx} \fi

\PrintedOrElectronic

\usepackage{t1enc,dfadobe}

\usepackage{egweblnk}
\usepackage{cite}

\usepackage{amsmath}
\newcommand{\norm}[1]{\left\lVert#1\right\rVert}
\DeclareMathOperator*{\argmin}{arg\,min}

\title{Self Functional Maps}

\author[O. Halimi \& R, Kimmel]
{\parbox{\textwidth}{\centering O. Halimi$^{1}$ and 
        R. Kimmel$^{2}$
        }
        \\
{\parbox{\textwidth}{\centering $^1$Faculty of Electrical Engineering, Technion University, Haifa 32000, Israel\\
         $^2$Faculty of Computer Science, Technion University, Haifa 32000, Israel
       } 
}
}

\begin{document}

\teaser{
 \includegraphics[width=0.9\linewidth]{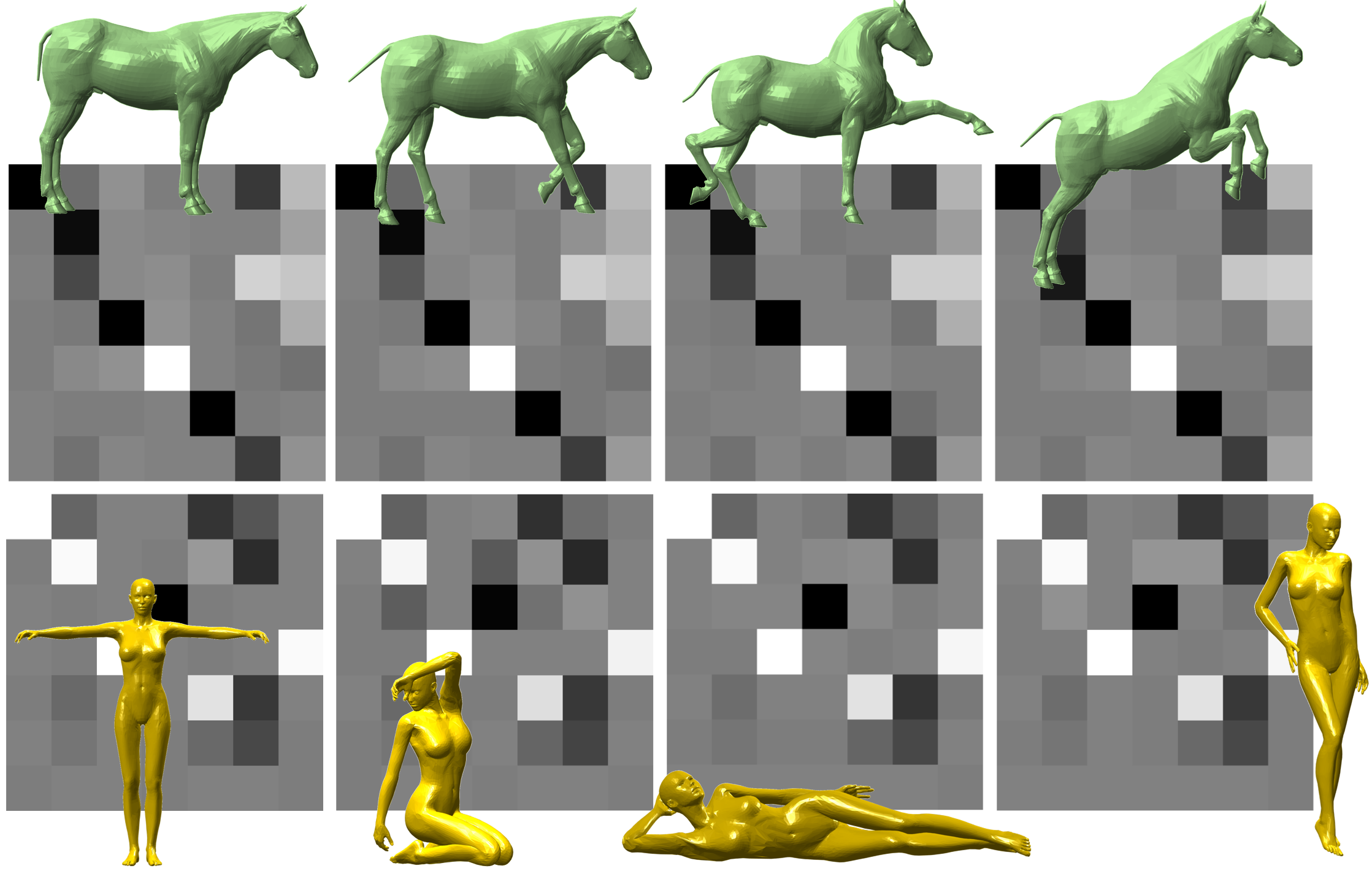}
 \centering
  \caption{{\em Self functional maps} of two classes of articulated objects at different poses. 
  The {\em self functional map} ($ 7\times 7$ matrix) of each shape is presented as a gray valued image behind its corresponding object.}
\label{fig:teaser}
}

\maketitle
\begin{abstract}
A classical approach for surface classification is to find a compact algebraic representation for each surface that would be similar for objects within the same class and preserve dissimilarities between classes.
We introduce {\em self functional maps} as a novel surface representation that satisfies these properties, translating the geometric problem of surface classification into an algebraic form of classifying matrices.
The proposed map transforms a given surface into a universal isometry invariant form defined by a unique matrix. 
The suggested representation is realized by applying the functional maps framework to map the surface into itself. 
The key idea is to use two different metric spaces of the same surface for which the functional map serves as a signature. 
Specifically, in this paper, we use the regular and the scale invariant surface laplacian operators to construct two families of eigenfunctions.
The result is a matrix that encodes the interaction between the eigenfunctions resulted from two different Riemannian manifolds of the same surface. 
Using this representation, geometric shape similarity is converted into algebraic distances between matrices.
\end{abstract}

\section{Introduction}

Since the scientific revolution attributed to Ren\'{e}  Descartes in the 15th and 16th centuries, researchers have been occupied with the problem of bridging between geometry and algebra.
With recent reincarnation of convolutional neural networks, some of the fundamental difficulties of translating geometric problems into algebraic ones are resurfacing.
In this paper we deal with the topic of computational surface classification.
Surfaces given as two dimensional curved manifolds are encountered in growing numbers on the Internet due to the rapid development of 3D sensing devices and 3D modeling techniques. 
Shape retrieval and classification problems are of great importance in the current age, when the prevalence of 3D modeling is increasing with acceleration in a variety of fields like e-commerce, virtual and augmented realities, computer aided design, 3D printing, medical 3D modeling, video games, the movies industry, and 3D mapping.
The rapid growth of the amount of geometric data requires the design of efficient content-based shape retrieval engines, that could perform similar tasks to those of existing search engines for textual information, using the shape itself as a query. 
For a proper operation of a geometry retrieval engine, the result should contain all occurrences of a query shape up to a predefined group of transformations. 
It is often desired to find a compact representation for a given shape that is invariant under transformations that keep the shape within its class. 

The properties of good signatures are compactness, structured representation, computational efficiency in extraction and retrieval, descriptiveness and robustness to deformations, and invariance to different parameterizations of the surface. 
The essence of a signature is to provide effective means to measure reliable similarity between shapes. 
A complete review on the topic of descriptors for 3D retrieval is beyond the scope of this paper and we refer the reader to \cite{li2014spatially} for a comprehensive survey of spectral shape descriptors for nonrigid 3D shape retrieval, and to \cite{liu2013survey} for a review with emphasis on partial shape retrieval. 
For a recent review on the problem of 3D shapes similarity assessment and the approaches for addressing it we refer to \cite{biasotti2016recent}.
Here, we provide a short incomplete sampling of recent efforts. 
In \cite{funkhouser2003search} an orientation invariant 3D shape-descriptor based on spherical harmonics was proposed. 
The requirement to pose invariance lead to use of LBO spectral geometry. 
In \cite{reuter2006laplace}, the authors proposed the use of Laplace Beltrami operator spectra as a ``Shape-DNA'' of surfaces and solids, achieving isometry invariance. 
The global point signature (GPS) was introduced in \cite{rustamov2007laplace}, where the histogram of pairwise distances in the GPS embedding space was used as a shape signature. 
In \cite{sun2009concise}, the heat kernel signature (HKS) was suggested and in \cite{dey2010persistent} a pose-oblivious matching algorithm was explored using the feature vectors calculated at the set of persistent HKS maxima. 
In \cite{bronstein2010scale} a scale-invariant version of the HKS was proposed. 
Isometry invariant volumetric descriptors based on a volumetric extension of the HKS were suggested in \cite{raviv2010volumetric}. 
Later on, the ``bag of geometric words'' approach for 3D shape retrieval was introduced by \cite{bronstein2011shape}, presenting Shape Google for isometry invariant shape retrieval.   
The same paradigm for partial shape retrieval was used in \cite{laga2011bag,lavoue2012combination}. 
Intrinsic Shape Context Descriptors (ISC) were introduced in \cite{kokkinos2012intrinsic}, generalizing shape context descriptors, previously used for analysis of planar shapes, to surfaces.   
 
Another perspective of the problem of measuring shapes similarity is treating each surface as a metric space and using the Gromov-Hausdorff distance between such geometric structures. 
The Gromov-Hausdorff framework for non-rigid shapes comparison was first suggested in \cite{memoli2005theoretical}. 
The same definition of the Gromov-Hausdorff distance \cite{burago2001course}  as the distortion of embedding one metric space into another, lead to the introduction of the generalized MDS framework (GMDS) for measuring non-rigid shapes similarity \cite{bronstein2006generalized}. 
In a follow up paper \cite{bronstein2010gromov}, the geodesic distances used in the Gromov-Hausdorff distance were replaced with diffusion distances. While in \cite{aflalo2016spectral} the GMDS problem was translated to the spectral domain. In \cite{berard1994embedding} the Riemannian manifold was embedded in the heat kernel space and Gromov-Hausdorff distance was measured in the heat kernel space.
In \cite{memoli2011gromov} the Gromov-Wasserstein distance was defined as a reformulation of the distance measure between the shapes. 

A different approach for quantifying shape similarity was to embed the given surfaces into a new space in which isometry is translated into simple transformations. 
One of the early papers that exploit this concept is \cite{schwartz1989numerical}, where surfaces were flattened to a plain while preserving their geodesic distances for the purpose of finding a unique parametrization of cortical surfaces in  computer aided neuroanatomy. 
Shapes similarity can be treated with conformal geometry tools, locally preserving the angles on the original surface. 
For example, in \cite{wang2007conformal} quasi-conformal maps for matching surfaces were explored.
Shape descriptors based on conformal geometry were proposed in \cite{ben2008characterizing}, using the conformal factor to quantify dissimilarity. 
Comparing shapes using conformal Wasserstein distance was suggested in \cite{lipman2011conformal}, measuring the optimal mass transportation distance between conformal factor profiles obtained by mapping surfaces to a unit disc. 
A different type of embedding is trying to preserve the metric properties of the original curved surface. 
Embedding of the surface into a low dimensional Euclidean space that translates the geodesic distances between surface points into distances in the Euclidean space was suggested in \cite{elad2003bending}, creating isometry invariant  signatures in the embedding space, referred to as {\em canonical forms}. 
Canonical forms were used for face recognition in \cite{bronstein2005three}. 
In \cite{ling2007shape} similar method was suggested for the classification of planar shapes.
In \cite{shamai2017geodesic} the Geodesic Distance Descriptor (GDD) embedding was proposed, mapping the original surface to the complex Euclidean space.
Diffusion maps \cite{coifman2006diffusion} can be regarded as yet another type of embedding into a Euclidean space, preserving the diffusion distances defined on the surface. 
Finally, in \cite{qiu2007clustering}, a graph embedding procedure was proposed, preserving the commute time between the nodes.
Here, we treat each surface as an interaction between two specific metric spaces, where the spectral analysis of such an interaction provides a useful representation for various geometry analysis and processing tasks. 

\section{Related work}
\subsection{Functional maps}
Functional maps \cite{ovsjanikov2012functional} were introduced as an approach for transferring functions between shapes
without using the explicit knowledge of point-to-point correspondence between the shapes. 
It provides a way to transform the representation of a function in the basis defined in the original shape to the representation of its mapped version in the basis of the destination shape. 
Due to their optimality in representing smooth functions \cite{brezis2017rigidity,aflalo2016best} the eigenfunctions of the Laplace-Beltrami operator (LBO) were chosen as a preferred basis.
Still, the method is not restricted to a specific choice of bases. 
The transformation between the representation coefficients is linear and can be represented by a matrix that can be calculated using a knowledge about the way known descriptors map from one shape to the other.
Here, we extend this concept and define the {\em self functional map} as a convenient universal representation of a given surface.
We will demonstrate the power of this representation as a shape signature for classification.

Functional maps have been extended in various directions that can be also applied to self functional maps. 
In \cite{nognengimproved}
it was suggested to extend the original basis with pointwise-product of the leading basis eigenfunctions. 
This method can be used to compactly and accurately compute functional maps using only the first few basis elements which are considered to be more stable than the higher frequencies extracted numerically from the LBO. Following the definitions in \cite{shtern2015spectral} the original basis can be enriched by adding the inner product of the spectral gradient fields, as well as the cross product between two spectral gradient fields in the normal direction. 
In \cite{rodola2017partial} it was suggested that there is an underlying relation between the functional maps of a shape and part of the shape. 
A similar analysis can be applied to relate between the self functional map of a shape as a whole and its partial version.


\subsection{Choice of a basis}
A surface coupled with a metric tensor defines a Riemannian manifold. 
The same surface can thus produce different Riemannian manifolds using different metrics.
The {\em self functional map} is defined as a functional map between two Riemannian manifolds generated from the same surface. 
Choosing a pair of metrics on the surface, the Laplace Beltrami operator is determined for each metric and the eigen-decomposition of the operators yields two different sets of basis functions. 
The {\em self functional map} transforms the representation coefficients of a scalar function between two different basis functions that are defined by the different metrics on the same surface.
The choice of the basis functions for the definition of a functional map is important. 
To obtain informative structures for off-diagonal entries of the self functional maps,  the basis functions should be derived from significantly different metrics.  
With an appropriate choice of metrics, the resulting self functional maps would yield unique algebraic structures that could serve for object classification. 
When the basis functions and the inner products are both isometry invariant, the resulting self functional map is isometry invariant by construction. 
In this paper the two specific metrics we have chosen to explore are the regular  and the scale invariant ones.
See, for example,  \cite{aflalo2013scale,raviv2011affine} for a scale and affine invariant metrics.

\section{Main contribution}
The suggested framework bridges between two apparently different methodologies.  
Namely, spectral transfer of functions between shapes, and the notion of shape invariant signatures.
We introduce a signature that reflects the interaction between two different metric spaces of the same manifold. 
The suggested {\em self functional maps} translate geometry analysis from its native mesh coordinates that lack the property of being universal, to an image processing like domain, with a regular structure and grid coordinates. 
The result is an efficient and compact representation, isometry invariant by construction, that could be extended to other groups of transformations beyond the isometry presented here. 

\section{Background}
In this paper, a shapes in 3D is considered as a surface identifying the shape with its boundary where, by definition, a surface is a two-dimensional manifold. 
Assigning a metric tensor $(g_{ij})$ to the surface $S$, it can be treated as a Riemannian manifold denoted as $\{S,g\}$. 
For the same surface, different metric tensors can be assigned like $g$ and $\tilde{g}$, resulting in different Riemannian manifolds like   $\{S,g\}$ and $\{S,\tilde{g}\}$, respectively. 
Given the Riemannian metric tensor, geometric quantities such as lengths of curves, distances on the manifold, angles and curvatures can be defined in terms of the metric tensor.  
Consider $S(u,v)$, a parametric surface, $S:\Omega \subset \mathbb{R}^2 \rightarrow \mathbb{R}^3$. 
Locally, the vectors $\{S_{u},S_{v}\}$ form a basis for the tangent plane $T_pS$, about each surface point $p \in S$. 
The {\em regular} metric tensor is defined as
\begin{equation}
g_{ij} \equiv \langle S_{\xi^i},S_{\xi^j}\rangle,
\end{equation}
where $\xi^1=u, \xi^2=v$. 
The metric is used to define an infinitesimal measure of length on the surface. 
Given a vector $w\in T_{p}S$ on the tangent plane defined at point $p \in S$, represented in the local coordinates of $\{S_u,S_v\}$, a small displacement about $p$ can be written as $dw = S_udu + S_vdv$, and an arclength on the surface is thereby defined as 
\begin{equation}
ds^2 = \langle dw,dw \rangle = g_{ij}d\xi^id\xi^j=(du \,\, dv )(g_{ij})
\left ( \begin{array}{l}du \cr  dv \end{array} \right ),
\end{equation}
 where we used Einstein summation convention  in the third term, and $(g_{ij})$ denotes the metric tensor in matrix form. 
Let $p_0,p_1 \in S$ be two points on the surface, and let $\Gamma(t) = S(u(t),v(t)),\, t \in [0,1]$ be a parameterized curve on the surface connecting these points so that $\Gamma(0) = p_0, \Gamma(1) = p_1$. 
Then, the length of the trajectory is given by summing the infinitesimal lengths along the curve
\begin{eqnarray}
  L(p_0,p_1,\Gamma) &=& \int_{p_0}^{p_1} ds = \int_0^1 |\Gamma'(t)|dt = \int_0^1 |S_uu_t + S_vv_t|dt   \cr 
   &=& \int_{u(0),v(0)}^{u(1),v(1)} |S_{u}du + S_{v}dv| \cr 
   &=&\int_{u(0),v(0)}^{u(1),v(1)} \sqrt{(du,dv)(g_{ij})(du,dv)^T}.
\end{eqnarray}
The distance between two points on the surface is calculated by taking the minimum of $L(p_0,p_1,\Gamma)$ over all  possible trajectories $\Gamma$ connecting the two points.

Two shapes are isometric if there exists a mapping from one surface to the other such that all the pairwise distances on the surfaces are preserved. 
Using the regular metric, isometric surfaces refer to transformations that are used to model different pose variations that semi-rigid bodies can undergo.
Therefore, shape signatures are expected to be isometry invariant.

In ~\cite{aflalo2013scale} a local scale-invariant pseudo-metric was defined as
\begin{equation}
	\tilde{g}_{ij} = |K|g_{ij}.
\end{equation}
Scale invariance in a differential terms was achieved by multiplying the regular metric by the local Gaussian curvature $K$. 
Intuitively, with this metric definition, the measure of interest is not the actual distance on the surface but rather the distance normalized by the geometric mean of the principal curvature radii. 
With this definition the pseudo-metric is both isometry-invariant as well as invariant under semi-local scaling of the shape. 
From the perspective of the scale invariant metric, regions of zero Gaussian curvature such as flat or cylindrical regions,  effectively shrink to a point. 
In order to make the scale invariant pseudo-metric a proper metric and resolve the degeneracy where the Gaussian curvature vanishes, Aflalo et al. slightly modified the metric to
\begin{equation}
	\tilde{g}_{ij} = \left (\sqrt{K^2+\epsilon^2 }\right )g_{ij},
\end{equation}
 for some small positive constant $\epsilon$.

\subsection{Similarity is metric dependent}
\begin{figure}[htb]
  \centering 
    \includegraphics[width=\linewidth]{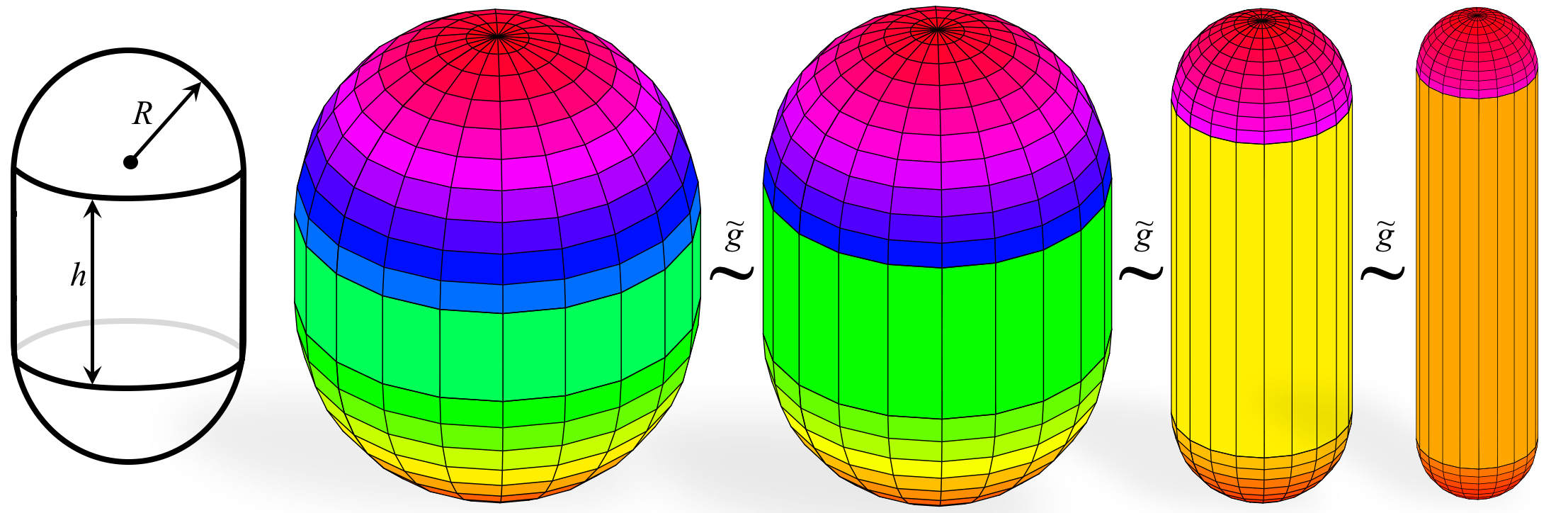}
  \caption{\label{fig:spherocylinder}
           spherocylinder. All the following spherocylinders are isometric with respect to the scale invariant metric.
           }
\end{figure}

Similarity between shapes can have a different interpretations depending on the metric one chooses to use. 
Two given shapes can be said to be isometric from a point of view of one metric and non-isometric from a point of view of a different one. 
To exemplify this statement, consider the two different Riemannian manifolds defined above, namely, 
the regular one defined by  $\{ S,g \}$, and the scale invariant one defined by $\{ S, \tilde{g}\}$. 
Considering a sphere of radius one, the pairwise distances on the surface overlap in both geometries, as the geodesic arc-length and the scale invariant arc-length coincide. 
Scaling the sphere by a uniform factor results in non-isometric surfaces from the regular perspective and isometric surfaces from the scale invariant one. 
Returning to the original sphere with radius one, assuming the sphere is cut symmetrically and the halves are glued with a thin stripe to form a spherocylinder with $R=1,\,\,h > 0$, as shown in Figure \ref{fig:spherocylinder}.
From the regular perspective, when $h\ll 1$, the deformed sphere is almost isometric to the original one as the introduction of the infinitesimal cylinder connecting the two half-spheres has almost no influence on the geodesic distances. 
In contrast, from the scale invariant point of view, this deformation has a dramatic effect, since on the cylinder part the Gaussian curvature vanishes.
The introduction of the infinitesimal cylinder leads to trajectories of distance that approach zero between every pair of points along the equator, effectively shrinking all the points along the equator to a single point. 
Next, assume the spherocylinder height $h$ is gradually extended.
As a consequence the geodesic distances between pairs of points on different halves of the sphere would increase when the regular metric is used, resulting in non-isometric family of surfaces.
While from the scale invariant point of view, all the resulting spherocylinders with varying $h$ are still isometric to one another. 
This example emphasizes the fact that similarity is influenced by the method used to measure distances on the surface, that is, by the definition of a metric tensor. 
Considering two different metric tensors can be thought of as two different observations of the same surface, where each is sensitive to different types of deformations. 

\subsection{Spectral geometries} 
\label{specg}
Shape analysis can be performed in the spectral domain, using the eigen-decomposition of the \textit{Laplace Beltrami operator}. 
Modeling the surface as a two dimensional Riemannian manifold $\{S,g\}$, possibly with boundary $\partial S$, the Laplace Beltrami operator (LBO) generalizes the classical Laplacian operator on the Riemannian manifold.
In local coordinates, the LBO can be expressed as
\begin{equation}
	\Delta_g \equiv \frac{1}{\sqrt{g}}\partial_i 
     						\sqrt{g}g^{ij}\partial_j, 
\end{equation}
 where $g^{ij}$ are the components of the inverse of the metric tensor $(g_{ij})$ and $g=\det(g_{ij})$  its determinant. 
The Laplace-Beltrami operator admits an eigendecomposition 
\begin{eqnarray}
	-\Delta_g\phi_i(s)  =& \lambda_i\phi_i(s) \quad & s\in S \setminus \partial S \cr 
    \langle\nabla_g \phi_i(s),\hat{n}(s)\rangle =& 0 \quad & s\in \partial S,
    \label{neumann}
\end{eqnarray}
with Neumann boundary condition (\ref{neumann}) for a surface with boundary, where $\hat{n}(s)$ is the normal to the surface along the boundary, and $\nabla_g$ is the intrinsic gradient defined on the Riemannian manifold.
The eigen-decomposition yields a discrete set of eigenfunctions that are invariant to isometries.
The eigenfunctions of the LBO on a regularly parametrized torus define the Fourier Transform, while its decomposition on general compact Riemannian manifolds produces eigenfunctions that, when ordered by their corresponding eigenvalues or frequencies, have been shown to be optimal for representing smooth functions on the manifold \cite{brezis2017rigidity,aflalo2016best}.
The invariance of the operator with respect to the isometry it was constructed by, makes it particularly useful for shape analysis. 
For example, pose variations can be modeled as isometries in their natural  sense $\{S,g\}$.
Thus, articulated shapes would share similar spectral properties with respect to $\Delta_g$.

For a comprehensive review of the literature in the field of spectral geometry we refer the reader to \cite{li2014spatially}. 
We provide here a short sampling of the literature. 
Diffusion maps \cite{coifman2006diffusion} have been used to embed a surface in the Euclidean space defined by the LBO eigenfunctions and eigenvalues. 
The Euclidean distance in the embedded space is equivalent to the diffusion distance on the surface.  In \cite{litman2011diffusion} introduced a diffusion-geometric framework for stable component detection in non-rigid 3D shapes, analogous to MSER in image analysis.
%
Heat Kernel Signatures (HKS) \cite{sun2009concise} were used as point descriptors defined by the LBO eigenfunctions and eigenvalues, that measure the rate of virtual heat dissipation from a surface point. 
It has been shown that HKS can be applied for detection of interest points on the surface as well as for shape matching and symmetry detection \cite{ovsjanikov2010one}. 
The Wave Kernel Signature (WKS) \cite{aubry2011wave} is a point descriptor that was introduced by solving the Schr\"odinger equation on the surface. 
Treating the solution as a quantum wave function that describes the position of some quantum particle, WKS represents its probability to be found in a specific point on the surface.
It can be defined by the LBO eigenfunctions and eigenvalues. 
The authors of \cite{aubry2011pose} demonstrated the usefulness of the descriptor for pose-consistent shape segmentation.
Global Point Signature (GPS) \cite{rustamov2007laplace} can be used for representation of the surface in the Euclidean space with coordinates defined using the eigenfunctions and eigenvalues of the LBO.
This embedding can be used for pose invariant segmentation and for shape classification using the histogram of pairwise distances between uniformly sampled surface points.

\begin{figure*}[htb]
  \centering
  \includegraphics[width=1\textwidth]{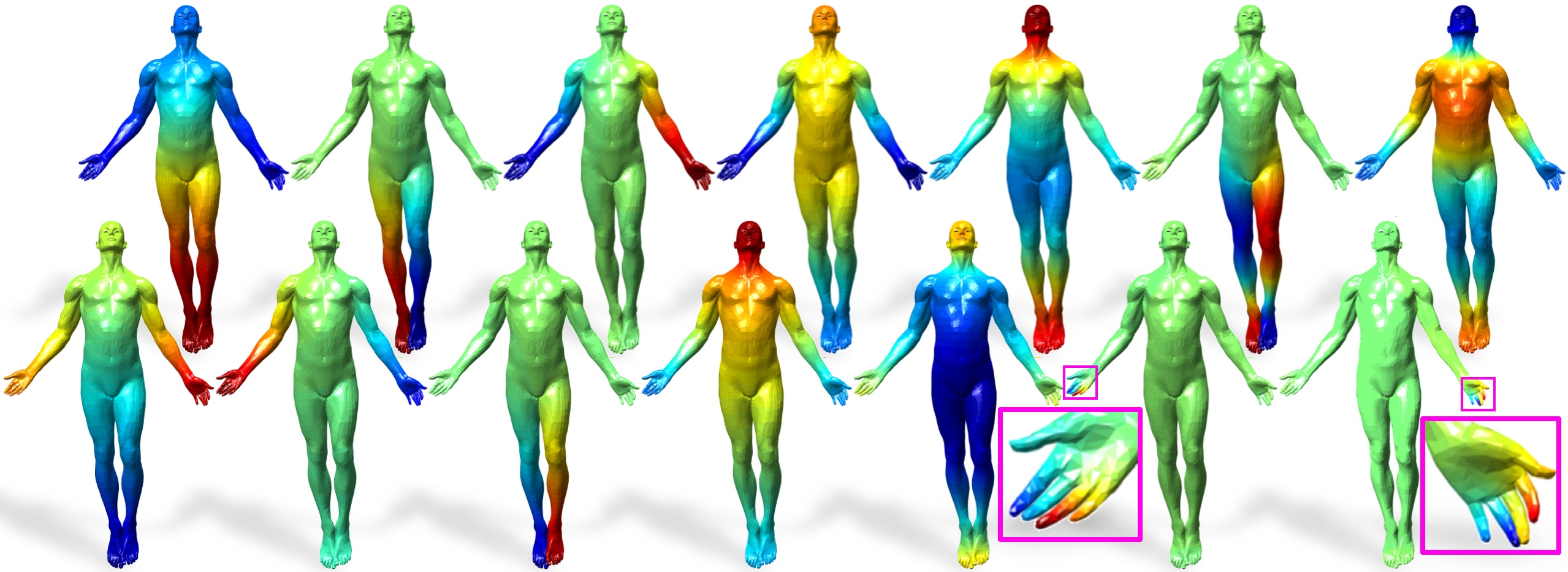}
  \caption{\label{fig:LBO}
Illustration of the first effective (excluding the constant) eigenfunctions of the regular (top) and scale invariant LBO (bottom). 
Color represents scalar values, where blue depicts low (negative) and red high (positive) values.
Zooming in on the phase transition regions of the scale invariant LBO's sixth and seventh eigenfunctions.
}
\end{figure*}

\begin{figure}[htb]
  \centering
  \includegraphics[width=0.34\linewidth]{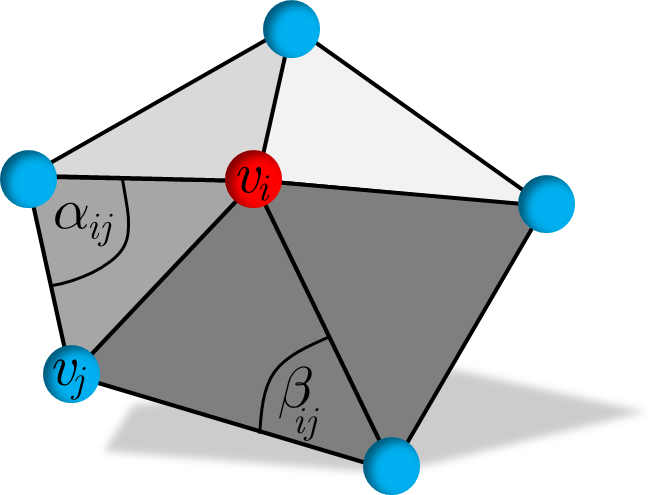}
  \caption{\label{fig:cot}
	Cotangent weights are derived from angles about each neighboring edge.}
\end{figure}

Defining the scale invariant Laplace Beltrami operator \cite{aflalo2013scale}, the conformal relation between the regular and the scale invariant metrics yields the following relation between the regular operator and the scale invariant one
\begin{equation}
\Delta_{\tilde{g}} = |K|^{-1}\Delta_g.
\end{equation}
The derivation of this nice property for surfaces is provided in  Appendix \ref{appen:conformal}.
The eigen-decomposition of the scale invariant LBO yields an isometry-invariant, differential scale-invariant discrete set of eigen-functions defined by
\begin{equation}
	-\Delta_{\tilde{g}}\tilde{\phi}_i = \tilde{\lambda}_i\tilde{\phi}_i. 
\end{equation}
Comparing the eigenfunctions of the operators, the scale-invariant LBO eigenfunctions are characterized by rapid changes of the phase in curved regions, while in flat regions the phase is approximately constant. 
For the regular LBO, the eigenfunctions 
 change of phase is observed in flat regions as well.
See Figure \ref{fig:LBO}. 
In the discrete domain, for triangulated surfaces the LBO is approximated using the cotangent weights scheme, see \cite{meyer2003discrete}, where 
\begin{equation}
L = A^{-1}W,
\end{equation}
 approximates the regular LBO $\Delta_g$, and 
\begin{equation}
\tilde{L} = K^{-1}A^{-1}W = K^{-1} L,
\end{equation}
is scale invariant version approximating $\Delta_{\tilde{g}}$.
Here, $W$ is the cotangent weights matrix.
It is defined by the angles $\alpha_{ij}, \beta_{ij}$ about the edge $ij$ that are depicted in Figure \ref{fig:cot}. 
Define $ \omega_{ij} = \frac{1}{2}(\cot{\alpha_{ij}}+\cot{\beta_{ij}}) $, then 
\begin{equation}
W_{ij} =\begin{cases} 
      \sum\limits_{j: (i,j)\in E} 	 \omega_{ij}  & i = j \cr
		-\omega_{ij}   & i \neq j,
   \end{cases}
\end{equation}
where $E$ is the set of all edges of our triangulated surface, and $A$ is a diagonal matrix of the per-vertex area.
That is, $A_{ii}$ is third the sum of areas of all the triangles containing vertex $i$. 
$K$ is a diagonal matrix, where $K_{ii}$ is the absolute value of the Gaussian curvature at vertex $i$, approximated, for example, by the angular deficiency formula, \cite{xu2009discrete}, and regularized by
\begin{equation}
	K_{ii} = \sqrt{\left ( \frac{2\pi-\sum_j\gamma_j^i}{A_{ii}}\right )^2 + \epsilon^2}
\end{equation}
 where $\gamma_{j}^{i}$ is the angle at vertex $i$ of the $j$-th triangle that contains vertex $i$, and $\epsilon$, as in the continuous case, is a small constant introduced to keep the metric valid when the Gaussian curvature vanishes.

\subsection{Functional maps between two surfaces}
Functional maps is a tool for transferring functions between surfaces without establishing an explicit point-to-point correspondence between the surfaces. 
Instead, the functional mapping is constructed using linear constrains, derived from partial knowledge about the mapping.
Let $S$, and $Q$ be two shapes, relating to each other by the transformation, $T:S \rightarrow Q$. 
Let $\mathcal{F}(S,\mathbb{R})$ be the real function space defined on $S$ and $\mathcal{F}(Q,\mathbb{R})$ be the real function space defined on $Q$.  
$T_F:\mathcal{F}(S,\mathbb{R}) \rightarrow \mathcal{F}(Q,\mathbb{R})$ maps every scalar function $f$ defined on $S$ to its corresponding function $g = f \circ T^{-1}$ defined on $Q$. $T_F$ is called \textit{the functional representation of the mapping $T$}. 
It has been shown that $T$ and $T_F$ can be recovered from each other. Clearly if $T_F$ is known, the vertex-correspondence can be calculated by mapping delta functions concentrated at each vertex from shape to the other.
Specifically, it has been shown that if the transformation $T$ is known, at least for some descriptor functions, the functional mapping can be constructed. 
Assume the orthonormal bases $\{\phi^S_i\}$ and $\{\phi^Q_i\}$ defined on each surface respectively. 
Suppose a function $f$ is defined on $S$, with the basis expansion
\begin{equation}
	f(s) = \sum\limits_i \alpha_i\phi^S_i(s) \quad s \in S.
\end{equation}
The mapped function $g =T_F(f)= f \circ T^{-1}$ on $Q$ can be expanded in the basis $\{\phi^Q_i\}$, as
\begin{equation}
	g(q) = \sum\limits_i \beta_i\phi^Q_i(q) \quad q \in Q,
\end{equation}
 and the relation between the expansion coefficients is linear and given by
\begin{equation}
	\beta_j = \sum\limits_i C_{ij}\alpha_i,
\end{equation}
where the matrix $C$ is the functional map matrix, given by
\begin{equation}
	C_{ij} = \langle T_F(\phi^S_i),\phi^Q_j\rangle.
\end{equation}
To solve for the matrix $C$, linear constrains are derived from the knowledge of specific corresponding functions on the two surfaces. 
Given a pair of corresponding functions $f:S \rightarrow \mathbb{R}$ and $g:Q \rightarrow \mathbb{R}$ with the coefficient vectors $\bar{\alpha}$ and $\bar{\beta}$ in the bases  $\{\phi^S_i\}$ and $\{\phi^Q_i\}$, respectively. 
The correspondence imposes the following linear constrain on $C$
\begin{equation}
	\bar{\beta} = C^{T}\bar{\alpha}.
\end{equation}

Corresponding functions are functions that preserve their value under the mapping $T$. 
For example, if $T$ is an isometry between the shapes, HKS or WKS signatures can serve as corresponding functions, or if a corresponding landmark or segment is given, the distance functions from it is corresponding between the shapes. 
Each pair of corresponding functions is translated to a linear constraint. 
Thus, the requirement for specific knowledge of the point-to-point correspondence is replaced by the relaxed requirement of knowledge about function correspondence.
In addition, other constraints can be imposed like commutativity with respect to specific operators, such as the Laplace Beltrami operator and symmetry operators.
Finally, the resulting optimization problem can be solved using numeric linear solvers.

\section{Functional maps between a surface to itself} 
Using two different metric tensors  $g$ and $\tilde{g}$, the same surface manifold $S$ can be regarded as two different Riemannian manifolds $\{S,g\}$ and $\{S,\tilde{g}\}$, respectively.
Next, we harness the functional maps formulation to define the {\em self functional map} as the functional map between the two different Riemannian manifolds $\{S,g\}$ and $\{S,\tilde{g}\}$. 
The transformation $T$ is defined as the trivial map $T: S \rightarrow S$ from the surface $S$ to itself, while the basis functions $\{\phi_i\},\{\tilde{\phi}_i\}$ are defined as the eigenfunctions of the corresponding Laplace Beltrami operators calculated with respect to the Riemannian metrics
\begin{eqnarray}
	-\Delta_g\phi_i &=& \lambda_i\phi_i \cr 
    -\Delta_{\tilde{g}}\tilde{\phi}_i &=& \tilde{\lambda}_i\tilde{\phi}_i.
\end{eqnarray}
The self functional map of the surface $S$ is defined as
\begin{equation}
C^{S,\tilde{S}}_{ij} \equiv \langle \phi^S_i,\tilde{\phi}^S_j\rangle,
\end{equation}
 where the inner product can be calculated with respect to $g$ or $\tilde{g}$ depending on the direction of the transformation. 
 The self functional map is a fundamental representation of the shape, reflecting the interaction between the eigenfunctions of the Laplace Beltrami operator calculated on different Riemannian manifolds originated in the surface that defined the shape. These metrics should be different yet intrinsic in the sense that distances can be computed without relating to the embedding space. One possible way to construct the self functional map is to use the regular metric tensor $g_{ij}$ and the scale invariant metric tensor $\tilde{g}_{ij} = |K|g_{ij}$.

The self functional maps are isometry invariant by construction since the eigenfunctions and the inner product are isometry invariant.
Therefore, for two isometric surfaces $S$ and $Q$, it holds that
\begin{equation}
	C^{Q,\tilde{Q}}_{ij} \equiv \langle \phi^Q_i,\tilde{\phi}^Q_j\rangle = \langle \phi^S_i,\tilde{\phi}^S_j\rangle \equiv C^{S,\tilde{S}}_{ij}.
\end{equation}

An interesting property of the self functional map is that it encodes the transformation of representation coefficients between basis functions, derived from different metrics, even between different isometric surfaces.  
Assume two isometric shapes $S$ and $Q$, and denote the LBO eigenfunctions calculated on the isometric surfaces $\{\phi^S_i\}$ and $\{\phi^Q_i\}$, respectively. Since the \textit{regular} Laplace Beltrami operator $\Delta_{g}$ is isometry invariant, the basis functions produced in this way are also isometry invariant. 
Assume the isometric transformation between the shapes is given by $T:S \rightarrow Q$, then, from the isometry invariance of the eigenfunctions it follows that
\begin{equation}
\label{phi}
T_F(\phi^S_i) \equiv \phi^S_i \circ T^{-1} = \phi^Q_i.
\end{equation}
In addition, assume we calculate on both shapes, $S$ and $Q$, a basis of the \textit{scale invariant} LBO eigenfunctions, denoted by $\{\tilde{\phi}^S_i\}$ and $\{\tilde{\phi}^Q_i\}$, respectively. 
Since the scale invariant Laplace Beltrami operator $\Delta_{\tilde{g}}$ is isometry invariant it follows that
\begin{equation}
T_F(\tilde{\phi}^S_i) \equiv \tilde{\phi}^S_i \circ T^{-1} = \tilde{\phi}^Q_i.
\end{equation}
Suppose we want to construct a functional map between $S$ and $Q$ and in $S$ we use the regular LBO eigenfunctions $\{\phi^S_i\}$ for representing functions, while in $Q$ we use the scale invariant LBO eigenfunctions $\{\tilde{\phi}^Q_i\}$.
Then, the functional map between $\{ S,g_S\}$ and $\{Q, \tilde{g}_Q\}$, denoted by $C^{S\tilde{Q}}_{ij}$, is given by
\begin{equation}
	\label{func_map}
	C^{S\tilde{Q}}_{ij} = \langle T_F(\phi^S_i),\tilde{\phi}^Q_j\rangle = \langle \phi^Q_i,\tilde{\phi}^Q_j\rangle = C^{Q\tilde{Q}}_{ij},
\end{equation}
 where the second equality follows from Equation (\ref{phi}). 
From this relation, since $C^{S\tilde{Q}}_{ij} = C^{Q\tilde{Q}}_{ij}$, it appears as if, in the case of isometric shapes, one can calculate the functional map between the shapes $S$ and $Q$, by operating only on the shape $Q$, without processing the shape $S$. 
In fact, in does not matter if the transformation is from $S$ to $Q$ or from $Q$ to itself, with respect to different metrics, we obtain the same linear transformation for interchanging the representation coefficients between the basis derived from the regular LBO and the basis derived from the scale invariant one. 
Therefore, the transformation depends only on the metrics by which the different basis functions are constructed and not on the source and destination surfaces of the transformation.
See Appendix \ref{appen:matching} for a different perspective on this idea. 

\section{Implementation considerations}
\subsection{Computing self functional maps}
Let $N$ denote the number of shapes in our dataset.
In order to calculate the self functional map for each shape $S_i,\,i \in \{1,\ldots,N\}$ we apply the following steps,
\begin{enumerate}
  \item The discrete regular Laplace Beltrami operator is calculated using the cotangent weights scheme
  \begin{equation}
  L = A^{-1}W.	
  \end{equation}
  The matrices are calculated according to the definition in Subsection \ref{specg}.
  \item The eigen-decomposition of the regular laplacian $L$ is computed and the eigenfunctions extracted up to order $n$,  and denoted by $\{\phi_k\}^n_{k=1}$
  \item The discrete scale invariant Laplace Beltrami operator is calculated using the expression
  \begin{equation}
  \tilde{L} = K^{-1}A^{-1}W = K^{-1}L,
  \end{equation}
  using the discretization in Subsection \ref{specg}.
\item The eigen-decomposition of $ \tilde{L}$ is calculated and the eigenfunctions  extracted up to order $m$, denoted by $\{\tilde{\phi}_k\}^m_{k=1}$.
\item The basis functions
$\{\phi_k\}^n_{k=1}, \{\tilde{\phi}_k\}^m_{k=1}$ 
 are obviously normalized with respect to the scale invariant metric,
\begin{eqnarray}
\phi_k^T KA\phi_k &=& 1 \quad \forall{k}\in\{1,\ldots,n\} \cr
\tilde{\phi}_k^{T}KA\tilde{\phi}_k &=& 1 \quad \forall{k}\in\{1,\ldots,m\} .
\end{eqnarray}
\item The self functional map $C \in \mathbb{R}^{m \times n}$ is calculated with respect to the scale invariant metric space,
\begin{equation}
   C_{pq} = \tilde{\phi}_p^T KA\phi_q.
\end{equation}
\item
To resolve the sign ambiguity of the eigenfunctions the self functional map was updated by Hadamard multiplication with a sign matrix $M\in \{-1,1\}^{m \times n}$
\begin{eqnarray}
	C_{ֿ\text{final}} &=& M \circ C \cr 
    M &=& ab^T, \,\,\,\,a \in \{-1,1\}^m , \,\,\,\,b \in \{-1,1\}^n,
\end{eqnarray}
where $a$ represents sign inversion of rows in the original matrix and $b$ represents sign inversion of columns. 
\end{enumerate}

In order to determine the sign vectors $a$ and $b$ for each shape, we follow the next procedure.
\begin{enumerate}
\item A representative shape of each class was chosen. 
Let $r:\{1,\ldots,N\} \rightarrow \{1,\ldots,N\}$ be a mapping from the index of the current shape $i$ to the index of the class representative $r(i)$. The self functional map of the representative shape $C_{r(i)}$ is kept unchanged with respect to the sign of its row and columns. The self functional map of the other shapes of the same class are updated by the sign matrix.
\item The sign vectors for the $i$'th shape are given by
\begin{equation}
a_i,b_i = \argmin_{\substack{a_i \in \{-1,1\}^m\\b_i \in \{-1,1\}^n}} \norm{C_{r(i)} - (a_{i}b^{T}_{i}) \circ C_i}_{L_1}.
\end{equation}
\end{enumerate}
\subsection{Classification}
The self functional maps $\{C_{i}\}^{N}_{i=1}$ can be used for shape classification. Classification was achieved by defining the following distance between the \textit{self functional maps}::
\begin{equation}
D_{ij} = \norm{C_i - C_j}^2_{L_1} \quad i,j \in \{1...N\}.
\end{equation}
With this distance the we could use a simple classification algorithm. Specifically we used k-means algorithm with $k = \#classes$.
Then the confusion matrix was calculated by using the true class labels and the labels predicted by the k-means algorithm.

Additionally, for visualization purpose, shape clustering can be performed by embedding the self functional maps into a Euclidean space $\mathbb{R}^d$ for low $d$, typically $d=2,3$.
This was done by the following procedure.
\begin{enumerate}
\item The distance matrix between the shapes, $D \in \mathbb{R}^{N \times N}_{+}$, is calculated using the pairwise squared $L_1$ distances between the self functional maps,
\begin{equation}
D_{ij} = \norm{C_i - C_j}^2_{L_1} \quad i,j \in \{1...N\}.
\end{equation}
\item The distance matrix is used to embed the shapes in the Euclidean space $\mathbb{R}^d$, given as an input to nonmetric MDS algorithm, using stress minimization.

\end{enumerate}
\section{Experimental results}
The experiments were done in Matlab 2017b, using the functions \textit{kmeans} and \textit{mdscale}. We ran \textit{kmeans} typically with few thousands iterations and few thousands replicates, and we used the squared $L_{1}$ distance between the matrices (not in Matlab interface), while \textit{mdscale} was used with the default configuration.

\begin{figure}[htb]
  \centering
  \includegraphics[width=\linewidth]{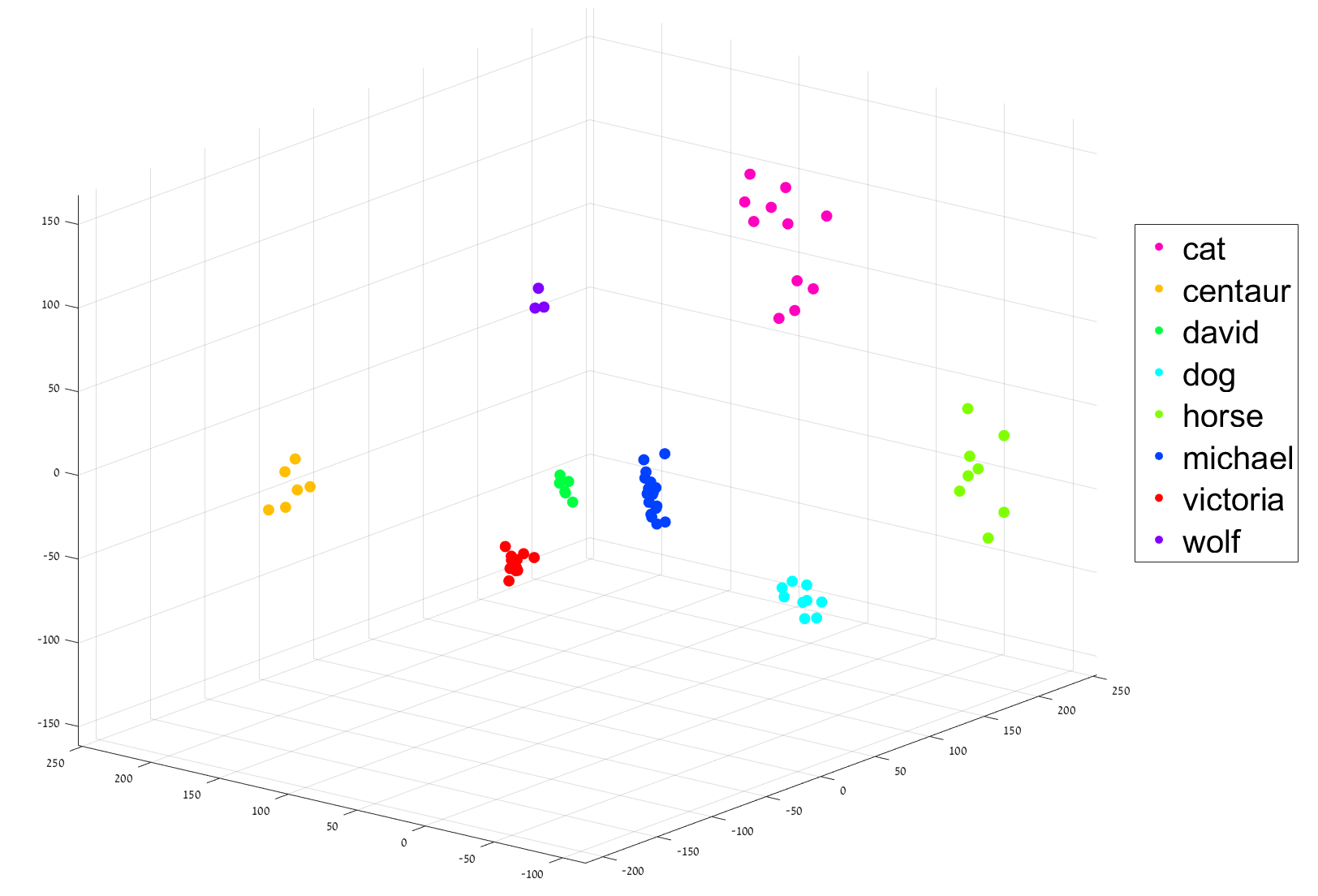}
  \caption{\label{fig:MDS_TOSCA}
           MDS clustering of the self functional maps in $\mathbb{R}^3$ calculated for the shapes in the TOSCA dataset, using $m=n=7$ eigenfunctions of the regular and the scale invariant LBOs. 
           Each point represents a shape. 
           Points of shapes within the same class have the same color.}
\end{figure}

We used TOSCA \cite{bronstein2008numerical} and SHREC 2010 \cite{bronstein2010shrec}, SHREC 2014 \cite{Pickup2014} and FAUST \cite{Bogo:CVPR:2014} datasets for our experiments. 
TOSCA dataset contains high-resolution ($\sim 50K$ vertices) 3D nonrigid shapes in a variety of poses. 
The database contains a total of $80$ objects, including cats, dogs, wolves, horses, centaurs, female figures, and two different male figures. 
The MDS clustering of the self functional maps in $\mathbb{R}^3$ is shown in Figure \ref{fig:MDS_TOSCA} with $n=7$ and  $m=7$ eigenfunctions of the regular and the scale invariant LBO, respectively.

\begin{figure}[htb]
  \includegraphics[width=0.7\linewidth]{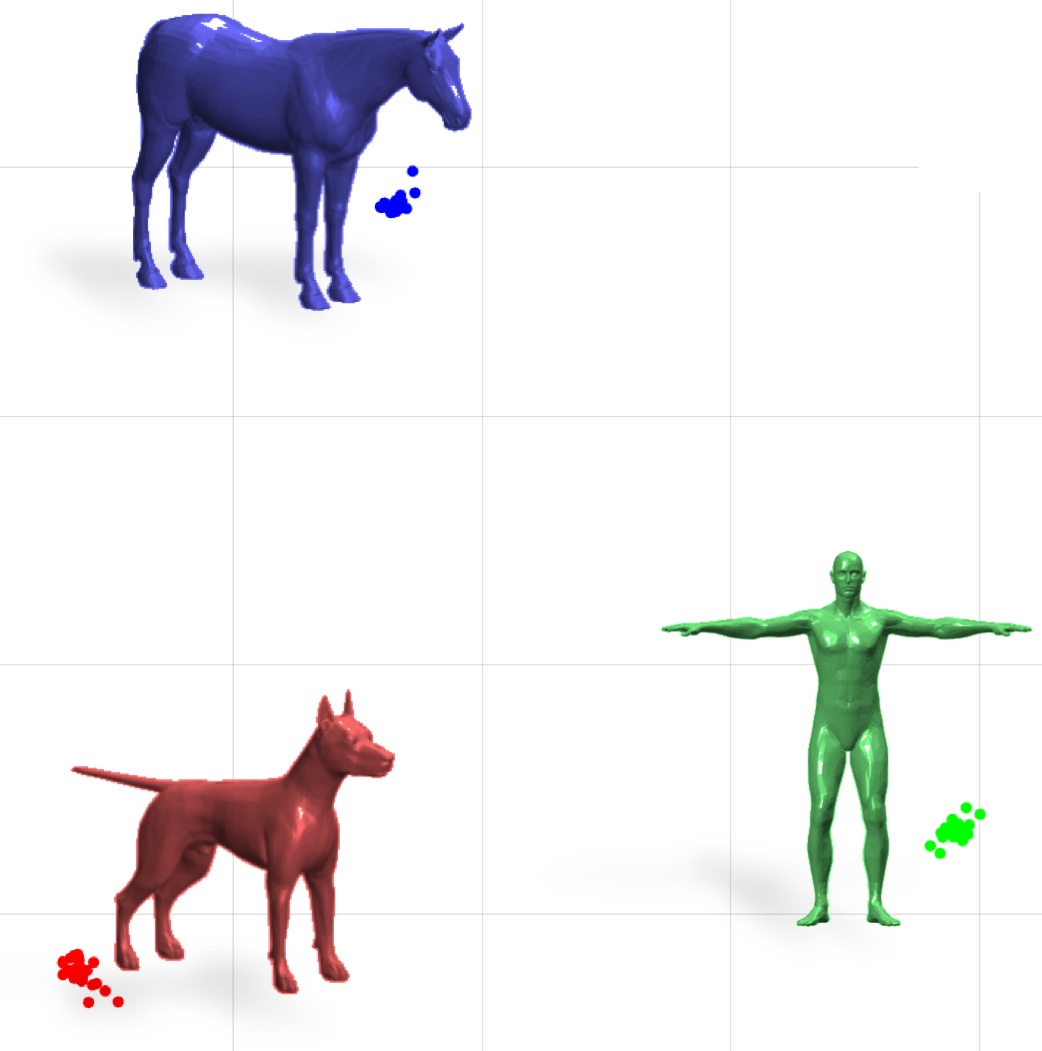}
  \centering
  \caption{\label{fig:mds_shrec}
           MDS display of the distance between the self functional maps in $\mathbb{R}^3$ computed for the SHREC'10 dataset, using $n=7$ eigenfunctions of the regular LBO and $m=7$ eigenfunctions of the scale invariant LBO.
         }
\end{figure}

In order to test the robustness of the self functional maps under a variety of deformations we used SHREC'10 dataset \cite{bronstein2010shrec}. 
The dataset consists of $138$ high-resolution ($10K-50K$ vertices) triangular meshes and contains three classes of objects, including human, horse and a dog, susceptible to non-rigid deformations as well as other distortions like noise, holes, topological changes, global and local scale, different sampling and more, as depicted by the following icons on the SHREC website \includegraphics[width=0.43\linewidth]{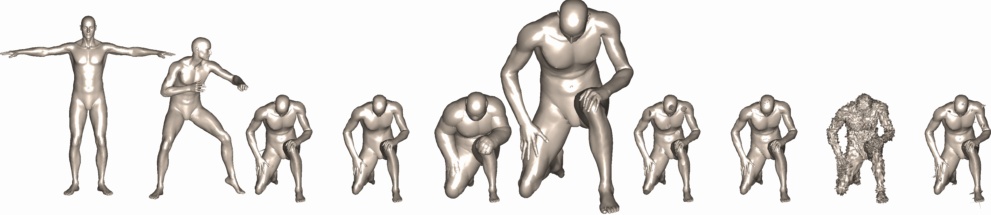}.
We demonstrate the performance of the self functional maps in classification in conditions of high intra-class variations. 
Figure \ref{fig:mds_shrec} shows the resulting MDS clustering of the self functional maps in $\mathbb{R}^3$ on SHREC'10, with $n= m=7$.


\begin{figure}[htb]
  \includegraphics[width=\linewidth]{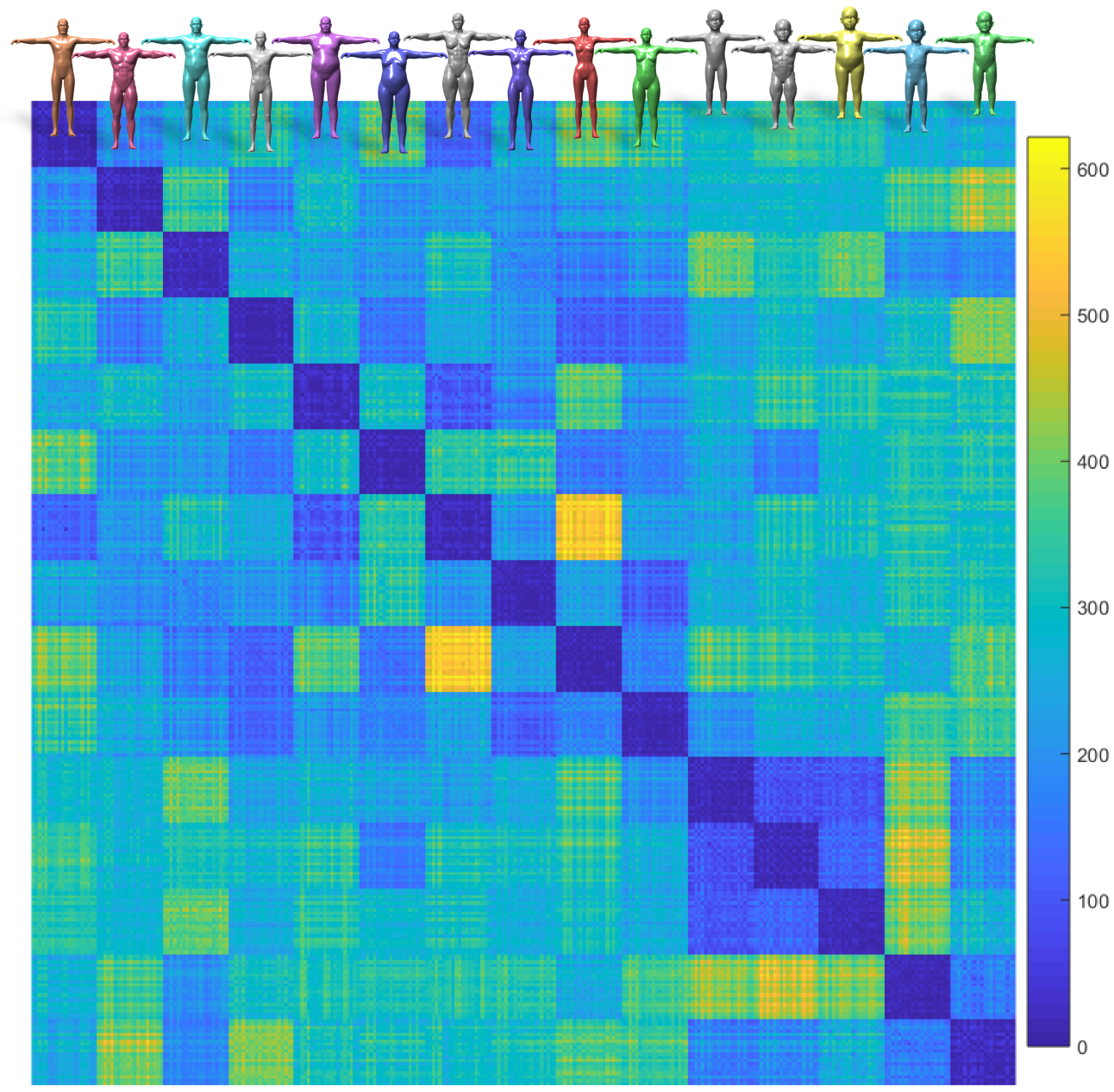}  
  \centering
  \caption{\label{fig:dist_shrec14}
Pairwise squared $L_1$ distance matrix between the {\em self functional maps}, calculated for the SHREC'14 synthetic data, with $m=n=8$ eigenfunctions for both the regular and the scale invariant LBOs. 
Bright colors represents large distances.
      }
\end{figure}

\begin{figure}[htb]
  \includegraphics[width=\linewidth]{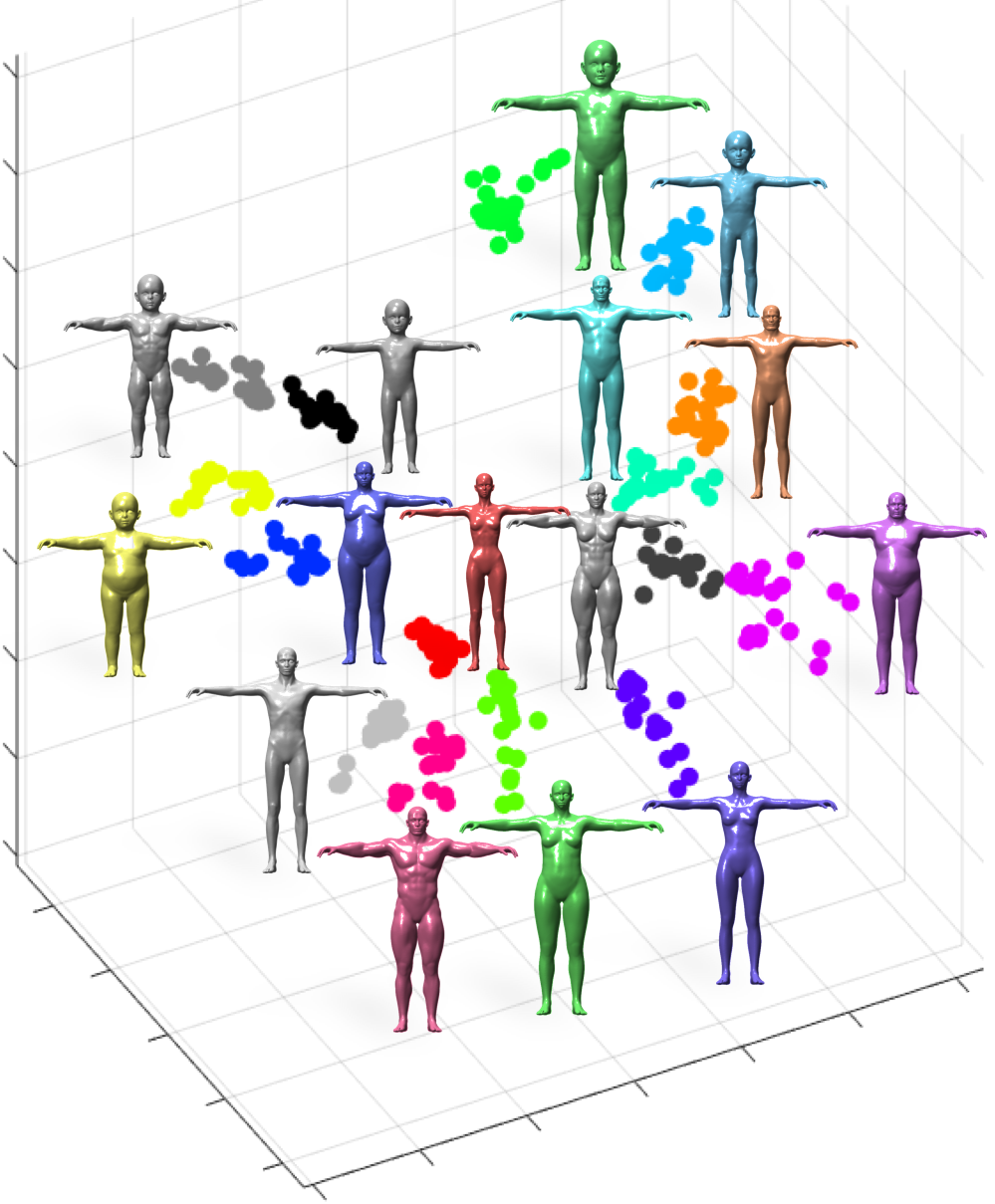}
  \includegraphics[width=\linewidth]{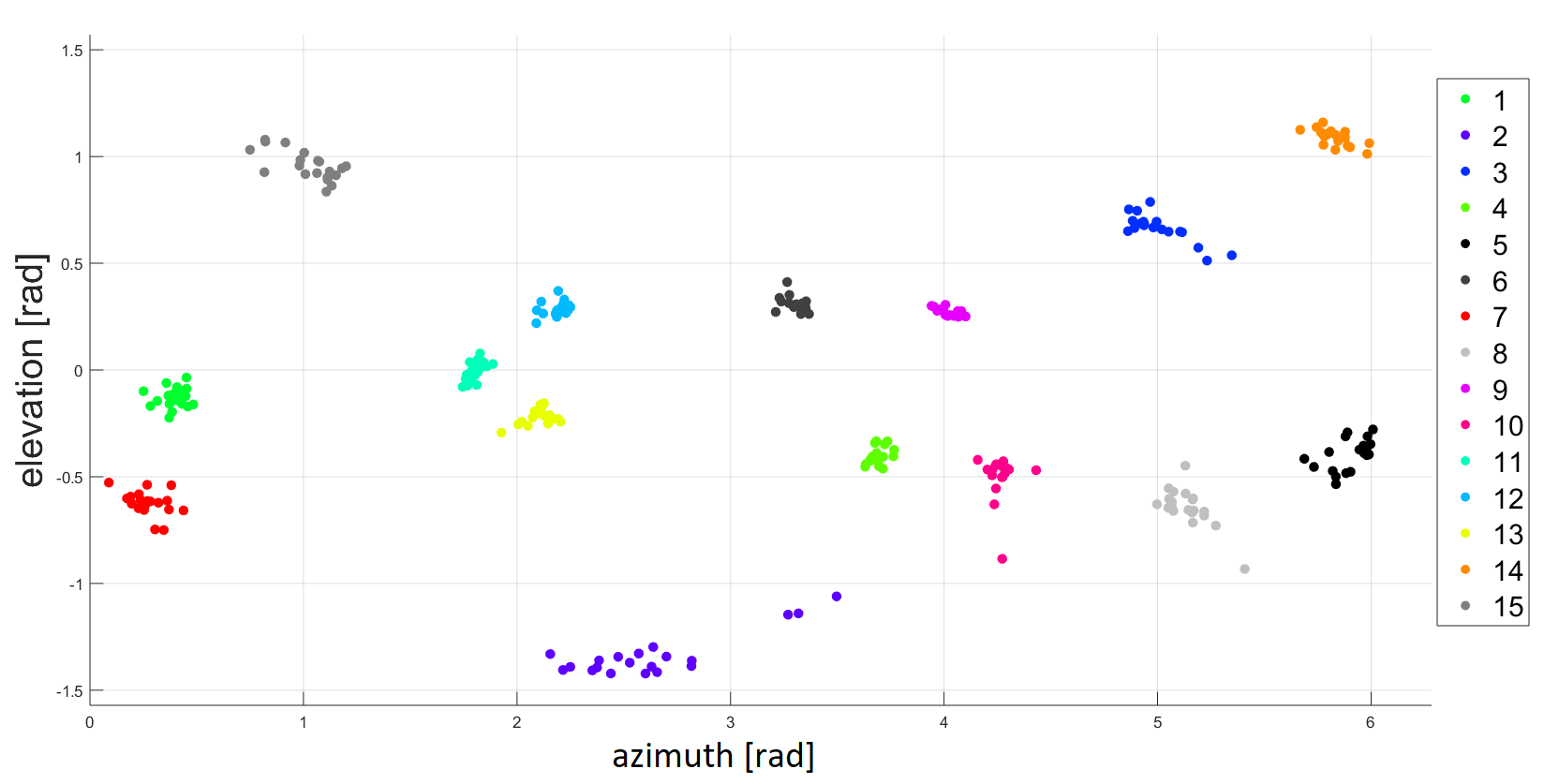}
  \centering
  \caption{\label{fig:mds_shrec14_syn}
           MDS embedding of the distance between the {\em self functional maps} into $\mathbb{R}^3$ computed for the SHREC'14 synthetic dataset, using $m=n=8$ eigenfunctions for both the regular and the scale invariant LBOs.
Each point represents a shape. Points of shapes within the same class have the same color. The clusters are well separated in 3D, for a better visualization the bottom figure shows the spherical angular coordinates of the points.
}
\end{figure}

To evaluate the efficiency of the {\em self functional maps} as signatures with the ability to separate between similar classes we conducted experiments on SHREC'14 dataset \cite{Pickup2014}.
This dataset consists of real and synthetic surfaces describing different human figures. 
The synthetic dataset consists of $15$ different human models, as depicted in the following icon taken from the SHREC'14 website \includegraphics[width=0.21\linewidth]{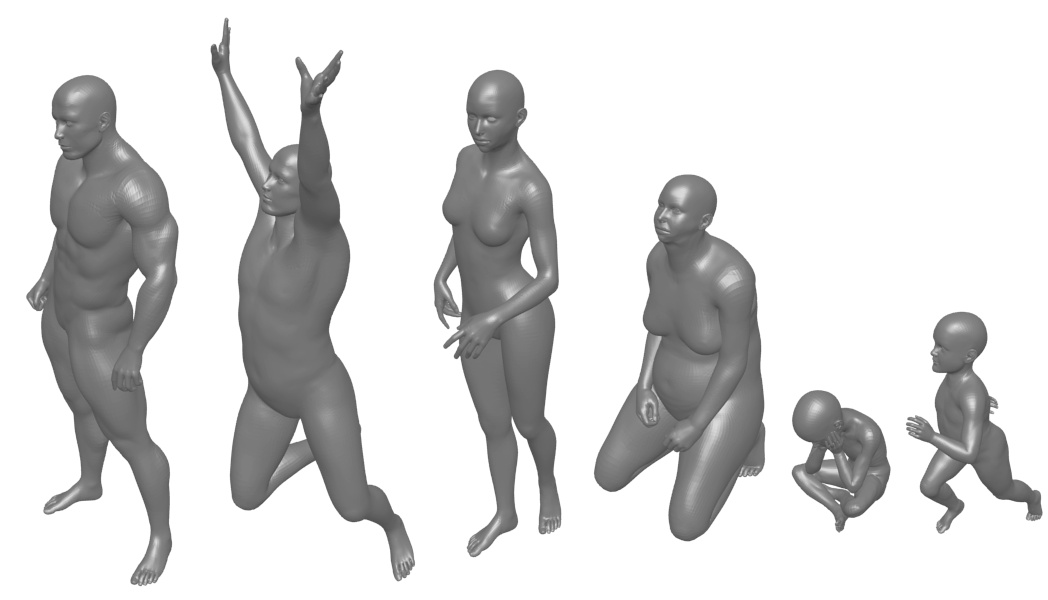},  each identity has its own unique body characteristics. 
Five of the figures are male, five female, and five are children. 
Each of these models appears in $20$ different poses, resulting in a dataset of $300$ models. 
The same poses have been applied to each virtual identity. 
A typical mesh contains around $60K$ vertices, and the meshes were down-sampled by a factor of $10$. 
Figure \ref{fig:dist_shrec14} shows the resulting pairwise squared $L_{1}$ distance matrix between the {\em self functional maps} of the different shapes, calculated on the SHREC'14 synthetic data, with $m=n=8$. 
The MDS embedding in $\mathbb{R}^3$ is depicted in Figure \ref{fig:mds_shrec14_syn}, in Cartesian coordinates and in spherical coordinates.

\begin{figure}[htb]
  \includegraphics[width=\linewidth]{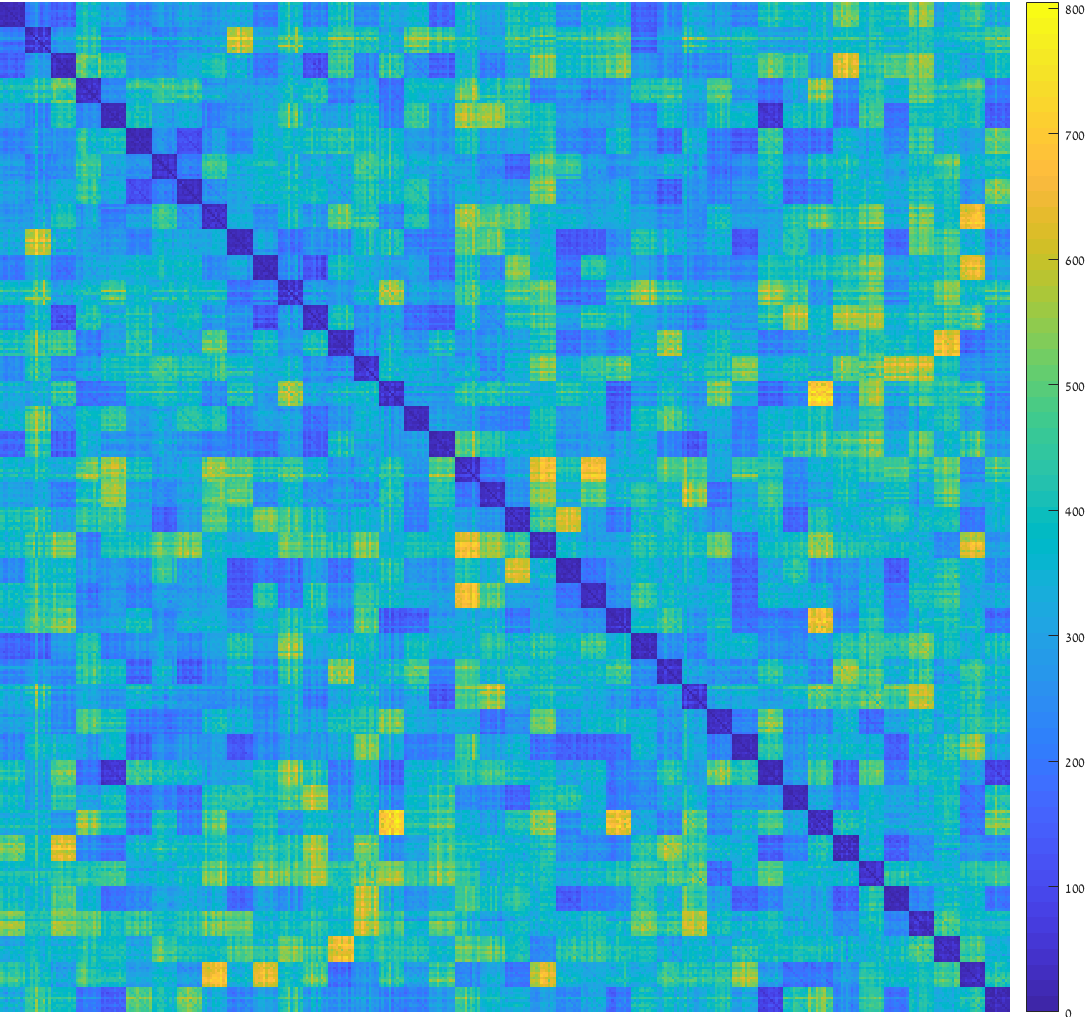}
  \centering
  \caption{\label{fig:dist_real}
Pairwise squared $L_{1}$ distance matrix between the {\em self functional maps}, calculated for the SHREC'14 real data, with $m=n=9$ eigenfunctions for both the regular and the scale invariant LBOs.
 Bright colors represents large distances.
        }
\end{figure}

\begin{figure}[htb]
  \includegraphics[width=\linewidth]{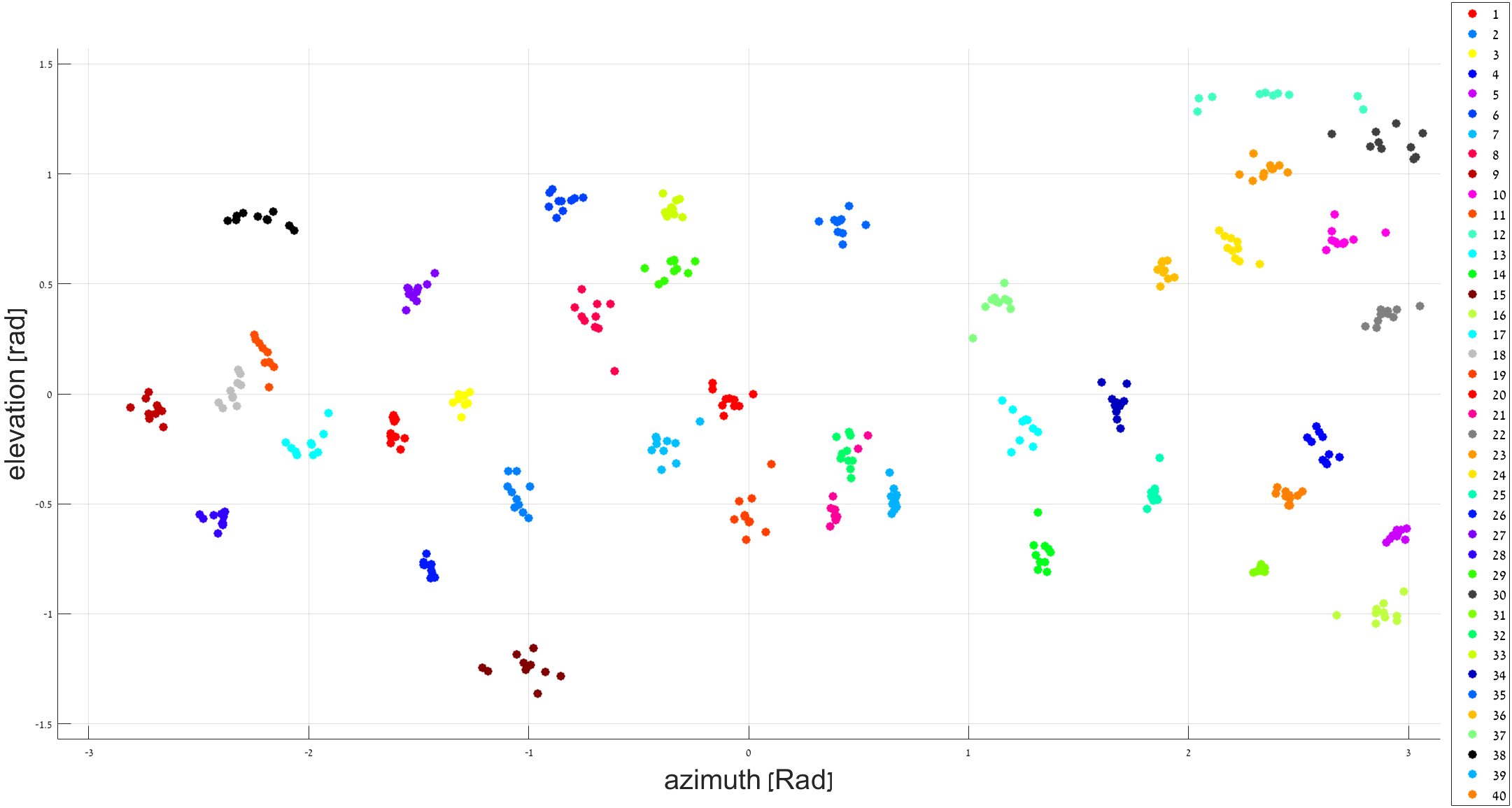}
  \centering
  \caption{\label{fig:mds_real}
           MDS embedding of the distances between the {\em self functional maps} into $\mathbb{R}^3$, computed for the SHREC'14 real dataset, using $m=n=9$ eigenfunctions of the regular  and the scale invariant LBOs. The points are displayed using spherical angular coordinates. Each point represents a shape. 
           Points of shapes within the same class have the same color. 
         }
\end{figure}

The real human dataset is composed of $400$ meshes, where there are $40$ human subjects, as depicted in the following icon taken from the SHREC'14 website \includegraphics[width=0.25\linewidth]{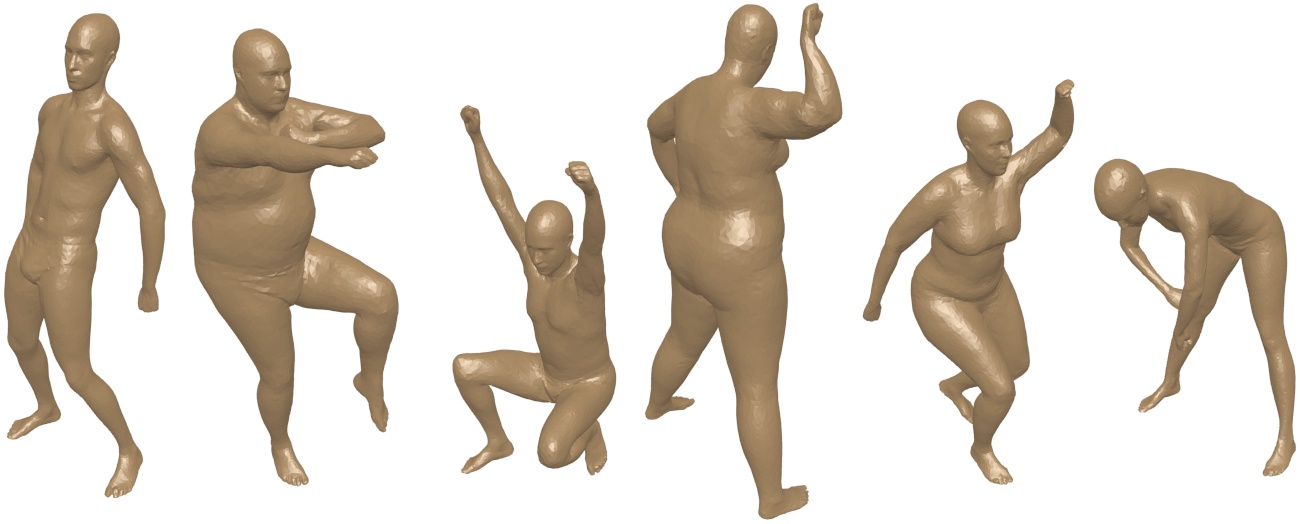}, each identity in $10$ different poses. 
Half the human subjects are male, and half are female. 
Here, again, the original meshes of  $\sim 15K$ vertices were down-sampled to $6000$ vertices.
The resulting pairwise squared $L_{1}$ distance matrix between the {\em self functional maps} of the different shapes, calculated on the SHREC'14 real data, with $m=n=9$ is depicted in Figure \ref{fig:dist_real}. The MDS embedding is depicted in Figure \ref{fig:mds_real} Since the real dataset contains a large number of classes and points, for a better visualization of the 3D embedding we display the points in spherical angular coordinates.

\begin{figure}[htb]
   \includegraphics[width=\linewidth]{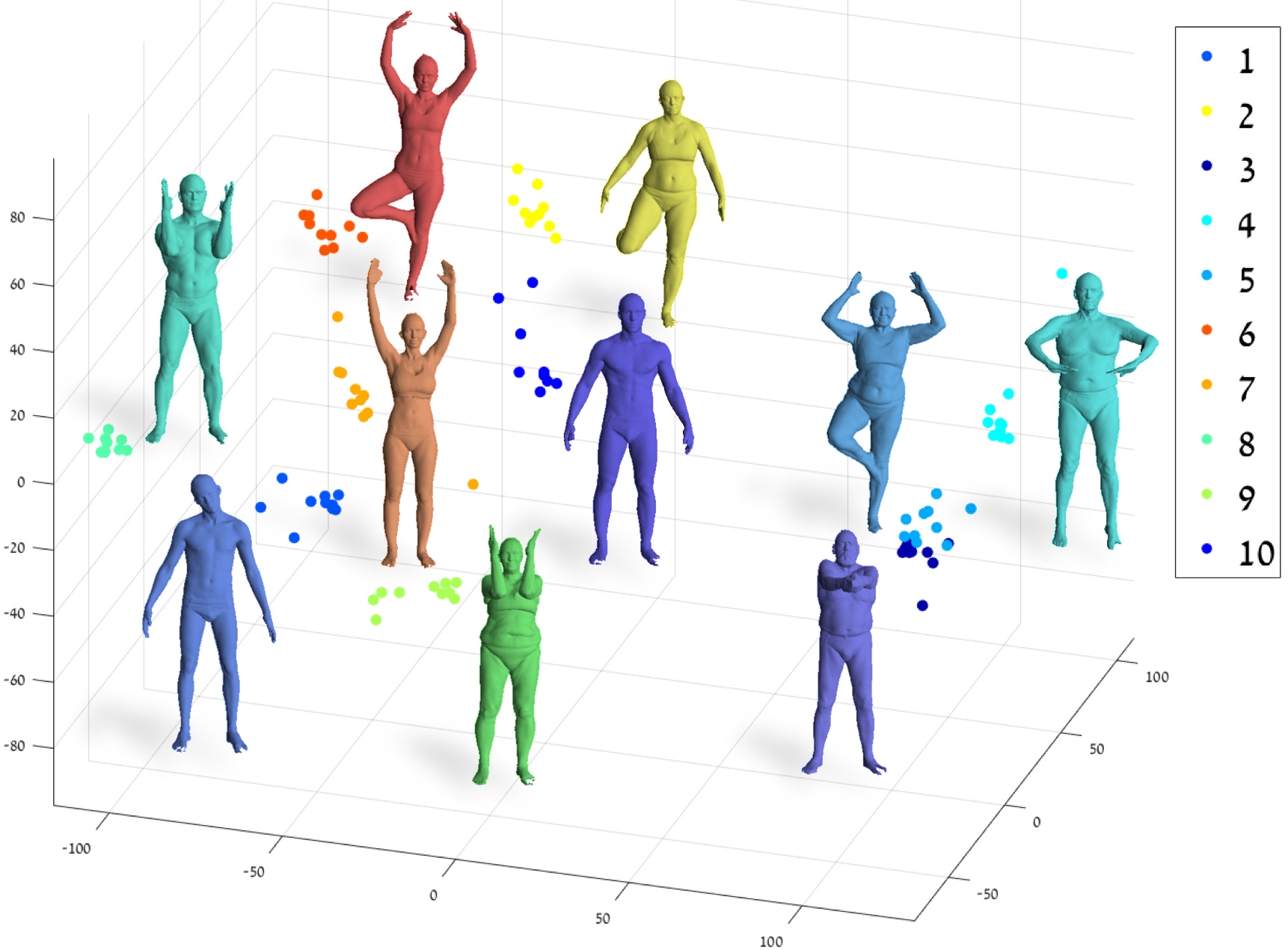}
 \centering
  \caption{\label{fig:mds_faust}
           MDS embedding of the distances between the {\em self functional maps} into $\mathbb{R}^3$, computed for the FAUST dataset, using $m=n=8$ eigenfunctions of the regular and the scale invariant LBOs. 
           Each point represents an identity in a specific pose. 
           Points of shapes within the same class (same identity) have the same color. 
         }
\end{figure}

In addition, we conducted experiments on the FAUST dataset \cite{Bogo:CVPR:2014} that contains real human scans of $10$ different human subjects, each subject appears in $10$ different poses, as depicted by the following icon from FAUST website \includegraphics[width=0.43\linewidth]{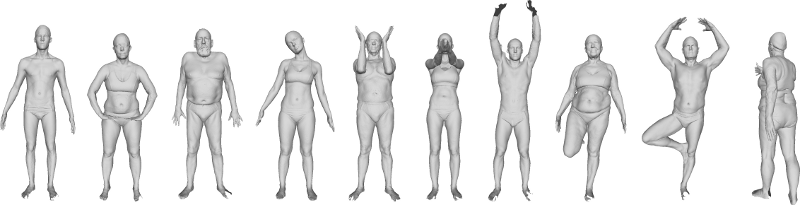}. 
Each mesh contains few thousands vertices. 
The MDS embedding is depicted in Figure \ref{fig:mds_faust}.

\begin{figure}[htbp]
    \includegraphics[width=\linewidth]{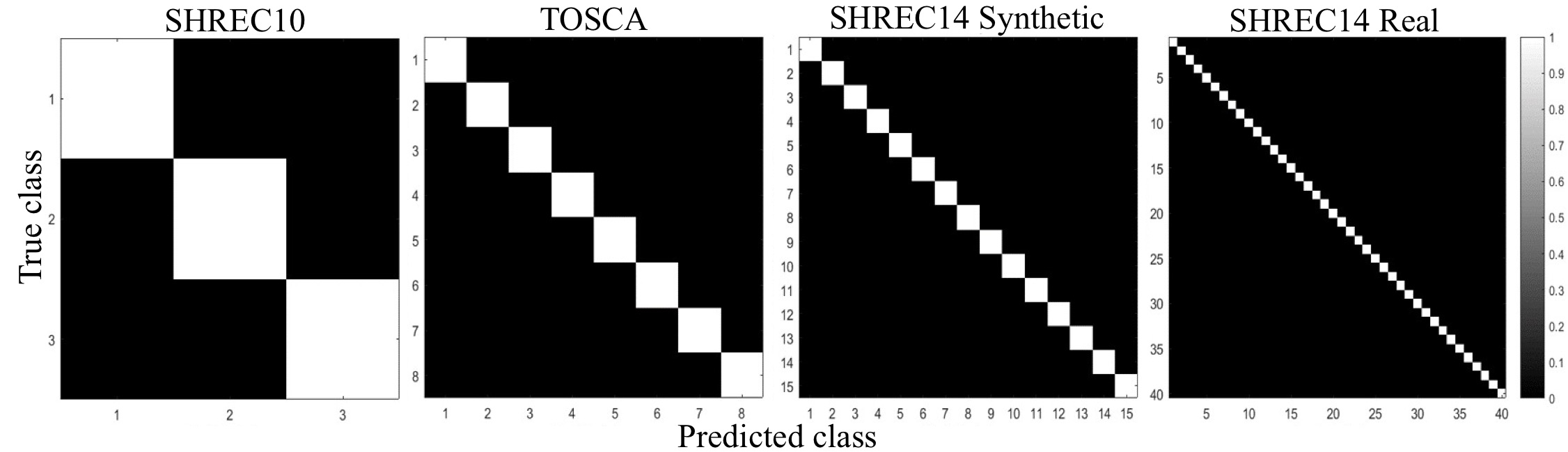}
  \centering
  \caption{\label{fig:confusion_maps}
Confusion matrices for each of the experiments. 
 Perfect classification is evident in all cases.
        }
\end{figure}

The confusion matrix was calculated for each dataset and is represented in Figure \ref{fig:confusion_maps}. Confusion matrix dimensions are $\#classes \times \#classes$. The vertical axis represents the \textit{true} class label and the horizontal axis represents the label predicted by the k-means algorithm. The value of the $ij$ entry is the number of shapes with \textit{true} class $i$ that were classified with label $j$, normalized by the \textit{total} number of shapes in class $i$ (values are between 0 and 1). The following confusion matrices show perfect classification the tested datasets. 
\section{Conclusion and future research directions}
The {\em self functional maps} framework was introduced as a universal representation for the task of classification of surfaces of articulated objects.
The suggested transition from a purely geometric problem into an algebraic form has the potential to be relevant for other applications in shape processing and analysis.
Possible examples are shape matching using either axiomatic classical methods or convolutional neural networks, in which the self functional map matrix serves as the input image. 
Note, that there is a meaningful order to the coordinates of the resulting matrix.
Other metrics could serve for generating alternative  self functional maps that are tailored for specific tasks. 
For example, in the case of matching between rigid man-made objects, different measures could be used to define different operators from which the signature could be derived. 
One example is the recent operator suggested in \cite{wang2017steklov}.
Stacking different metric pairs we could define a {\em self functional tensor}, as a tensor of {\em self functional maps} derived from different pairs of operators.
Future work can conduct perturbation analysis, similar to the one used in \cite{rodola2017partial}, on the {\em self functional map} to discover the influence of partiality on the maps. 
Extending the original set of the eigenfunctions by using pointwise products
 or local products of their gradients, similar to those presented in \cite{shtern2015spectral,nognengimproved},  it is possible to create larger self functional maps matrices containing additional information which is supported by reliable high frequency functions. 

\bibliographystyle{eg-alpha-doi}

\bibliography{egbibsample}

\newpage
\appendix
\section{From classification to matching}
\label{appen:matching}
Assuming we have classified two surfaces $S$ and $Q$ to belong to the same class.
Then, we have that their self-functional maps are similar to one another. 
That is, for all $i$ and $j$ we have
\begin{eqnarray}
 C^{S,\tilde{S}}_{ij} = \langle \phi_i^S,\tilde{\phi}_j^S \rangle \approx
 \langle \phi_i^Q,\tilde{\phi}_j^Q \rangle =  C^{Q,\tilde{Q}}_{ij} .
\end{eqnarray}
Besides our ability to use this property to resolve the signs ambiguity, we also have, as a by-product a way to match the points on the surface $S$ to those of the surface $Q$.  
That is, we would like to find the permutation $P$, such that for all $i$ we have $\tilde{\phi}_i^Q = P\tilde{\phi}_i^S$.
Again,  $P:S\rightarrow Q$ is a permutation matrix or matching function we are looking for. 

Next, we will prove that 
 \begin{eqnarray}
 \langle \phi_i^S,\tilde{\phi}_j^S \rangle \approx
 \langle \phi_i^Q,P\tilde{\phi}_j^S \rangle.
\end{eqnarray}
We can write $P$ as 
\begin{eqnarray}
 P =  \Phi^Q \mathbb{\rho} (\tilde{\Phi}^S)^T,
\end{eqnarray}
 where the columns of $\Phi^Q$ are given by $\phi_i^Q$, and similarly for $\tilde{\Phi}^S$,
  and $\rho$ is a connection matrix that we argue to be trivially given by $C^{S,\tilde{S}}$, [cite SGMDS].
  
Let us assume, without loss of generality, that 
\begin{eqnarray}
 \langle \phi_i^S,\tilde{\phi}_j^S \rangle =
 \langle \phi_i^Q,\tilde{\phi}_j^Q \rangle.
\end{eqnarray}
Then,
\begin{eqnarray}
P\tilde{\phi}_j^S &=&  \Phi^Q \mathbb{\rho} (\tilde{\Phi}^S)^T \tilde{\phi}_j^S = 
 \Phi^Q \mathbb{\rho} \delta_j^S,
\end{eqnarray}
 where $\delta_j$ is a vector of zeros with one at the $j$th entry. 
We readily have that 
 \begin{eqnarray}
  \langle \phi_i^Q,P\tilde{\phi}_j^S  \rangle 
     &=&  (\phi_i^Q)^T \Phi^Q \mathbb{\rho} \delta_j^S =
     (\delta^Q_i)^T \mathbb{\rho} \delta_j^S = \rho_{ij}.
\end{eqnarray}
Thus, recalling $P = \Phi^Q \mathbb{\rho} (\tilde{\Phi}^S)^T$, and that 
 $\rho_{ij} = \langle \phi_i^S,\tilde{\phi}_j^S \rangle$
 we conclude that given the self-functional maps of two surfaces that belong to the same class we could easily obtain a functional map between the two surfaces. 

\section{Conformal maps \& scale invariant LBO for surfaces}
\label{appen:conformal}
Given a Riemmanian manifold $\{S,g\}$ of dimension $n=2$, and let 
\begin{equation}
	\tilde{g}_{ij} = \mu g_{ij},
\end{equation}
 be a conformal mapping of the metric $g_{ij}$, where $\mu\in \mathbb{R}^+$ is the conformal factor. 
Recall the LBO definition,  
\begin{equation}
    	\Delta_g = \frac{1}{\sqrt{g}}\partial_i \sqrt{g}g^{ij}\partial_j.
\end{equation}
Taking the inverse of the metric tensor we obtain 
\begin{equation}
	\tilde{g}^{ij} = \mu^{-1} g^{ij}.
\end{equation}
Denoting the determinant of the metric by $g$ and the determinant of its conformal mapping by $\tilde{g}$, we readily have that 
\begin{eqnarray}
	       \tilde{g} &=& \mu^2 g \cr
    \sqrt{\tilde{g}} &=& \mu \sqrt{g}.
\end{eqnarray}
Therefore, the LBO of a conformal scaled metric is given by
\begin{eqnarray}
   \Delta_{\tilde{g}} &=& \frac{1}{\sqrt{ \tilde{g}}}\partial_i \sqrt{\tilde{g}}\tilde{g}^{ij}\partial_j \cr
    &=& \frac{1}{\mu\sqrt{g}}\partial_i \mu\sqrt{g} \mu^{-1} g^{ij}\partial_j \cr
    &=& \frac{1}{\mu \sqrt{g}}\partial_i \sqrt{g}g^{ij}\partial_j \cr
    &=& \mu^{-1} \Delta_{g}. 
\end{eqnarray}

\end{document}